\documentclass[twocolumn,showpacs,amsmath,amssymb,floatfix,prl,superscriptaddress]{revtex4-1}

\usepackage{graphicx}
\usepackage{xspace}
\usepackage{siunitx}
\usepackage[usenames]{color}
\usepackage{comment}
\usepackage{bigdelim}
\usepackage{amssymb}
\usepackage{amsmath}
\usepackage{array}
\usepackage{multirow}
\usepackage{tabularx}
\usepackage{dcolumn}
\usepackage{booktabs,dcolumn}
\usepackage{bm}
\usepackage{float}
\usepackage{fancyhdr}
\usepackage{afterpage}
\usepackage{url}
\usepackage{bm}
\usepackage{epstopdf}

\usepackage{enumitem}
\usepackage{setspace}
\usepackage{color}
\usepackage{hyperref}
\hypersetup{colorlinks,urlcolor=blue,citecolor=black}
\usepackage{listings}
\usepackage{geometry}
\usepackage[dvipsnames]{xcolor}

\usepackage[utf8]{inputenc} 
\usepackage{subfigure}
\usepackage{ulem}

\graphicspath{{figures/}}

\newcommand{\nba}[1]{} 
\newcommand{\rw}{$R_w$} 

\begin{document}

\title{Towards scanning nanostructure x-ray microscopy}

\author{Anton Kovyakh}
\affiliation{Niels Bohr Institute, University of Copenhagen, Universitetsparken 5, DK-2100 Copenhagen, Denmark}

\author{Soham Banerjee}
\author{Chia-Hao Liu}
\author{Christopher J. Wright}
\affiliation{Department of Applied Physics and Applied Mathematics, Columbia University, New York, NY 10027}

\author{Yuguang C. Li}
\affiliation{Department of Chemistry, University at Buffalo, The State University of New York, Buffalo, New York, NY 14260}
\author{Thomas E. Mallouk}
\affiliation{Department of Chemistry, The University of Pennsylvania, Philadelphia, PA, USA}

\author{Robert Feidenhans'l}
\affiliation{Niels Bohr Institute, University of Copenhagen, Universitetsparken 5, DK-2100 Copenhagen, Denmark}
\affiliation{European XFEL, D-22869 Schenefeld, Germany}

\author{Simon J. L. Billinge}
\affiliation{Department of Applied Physics and Applied Mathematics, Columbia University, New York, NY 10027}
\affiliation{Condensed Matter Physics and Materials Science Department, Brookhaven National Laboratory, Upton, NY 11973}

\date{\today}

\begin{abstract}
We demonstrate spatial mapping of the local and nano-scale structure of thin film objects using spatially resolved PDF analysis of synchrotron x-ray diffraction data.
This is demonstrated in a lab-on-chip combinatorial array of sample spots containing catalytically interesting nanoparticles deposited from liquid precursors using an ink-jet liquid handling system. We present a software implementation of the whole protocol including an approach for automated data acquisition and analyis using the atomic pair distribution function (PDF) method. The protocol software can handle semi-automated data reduction, normalization and modelling, with user-defined recipes generating a comprehensive collection of metadata and analysis results. By slicing the collection using included functions it was possible to build images of the 2D object containing using different quantities for contrast, allowing us to determine the spatial map relating to different aspects of the local structure on the array.
\end{abstract}

\maketitle

\section{Introduction}

The nanoscale structure of a material has a critical impact on the properties~\cite{billi;s07}.  Important nanostructured materials can consist of discrete nanoparticles~\cite{banerjeeClusterminingApproachDetermining2020h}, short-range nanostructural modifications to another well-ordered structure~\cite{bozinLocalOrbitalDegeneracy2019f}, nanoporous structures~\cite{jacksonHomoLigandMixedLigandMonolayerProtected2006} and so on.  In the past few decades structural tools have emerged for studying nanostructure.  When it is static, imaging methods such as transmission electron microscopy (TEM) and scanning tunneling microscopy (STM) can yield direct images of nanostructural features~\cite{jacksonHomoLigandMixedLigandMonolayerProtected2006,turnerDirectImagingLoaded2008}. On the other hand, diffraction methods such as the atomic pair distribution function analysis of powder~\cite{egami;b;utbp12} or single crystal~\cite{weberThreedimensionalPairDistribution2012d} data yield quantitative nanostructural information~\cite{egami;b;utbp12} from static and fluctuating nanostructures.  For larger objects it can be desirable to combine spatially resolved (microscopic/imaging) approaches with diffraction to elucidate static, spatial variations in local nanostructure.  For example, this approach was demonstrated by combining PDF analysis with computed tomography (ctPDF)~\cite{jacqu;nc13}, giving spatial maps of local nanostructure of slices through bulk objects such as spiral wound AA batteries~\cite{jensenXRayDiffractionComputed2015}

An important sample geometry is that of a thin-film on a substrate.  Here we explore making spatially resolved nanostructure maps from nanostructured samples on a thin substrate. This is made possible by the recent demonstration that reliable PDFs could be obtained from nanostructured films in normal incidence (tfPDF)~\cite{jense;ij15}, combined with rapid scanning that is at the heart of the ctPDF development. This could be used for example, for the
analysis of combinatorial arrays of thin film libraries on a chip, a synthesis method which has become a widely accepted industry standard\cite{Service:1997wd,XIANG:1995tn} in many fields including heterogeneous catalysis, pharmaceuticals, biomaterials, optics and multi-principal element alloys\cite{Senkan:1998wx,Daly:2015eb,Kohn:2004ju,Chan:2015kz,Miracle:2017bi,Potyrailo:2011jo}. We describe here a proof of principle experiment along with python scripts that can be used to handle such spatially resolved data. This shows that high-throughput scanning probes of nanostructure are possible in thin film geometry resulting in images of the spatial distribution of different nanostructure parameters such as lattice parameters, atomic positions
and atomic displacement parameters, nano-crystallite size and so on.

In the lab-on-a-chip experiment one of the key steps is to relate positional information (where the beam hits the sample) with measured data in the form of diffraction images and any prior information from the sample preparation such as target composition.
Automation is a priority at modern x-ray synchrotron beamlines where metadata about the instrument configuration, such as motor positions, is available electronically.

Here we describe a protocol for handling this type of analysis, including data acquisition at the XPD powder diffraction instrument at NSLS-II, data reduction that tolerates sample heterogeneity, and subsequent data analysis using the pair distribution function (PDF) technique.
The accompanying software allows the data to be reduced and analyzed in a highly automated fashion, and the extracted material specific properties to be easily visualized as 2D parameter maps.

As a demonstration we consider an array of catalytic nanoparticles on a carbon paper substrate using an ink-jet printing approach to allow for deposition of hundreds to thousands of distinct compositions of nanoparticles on a single substrate~\cite{Reddington:1998wc}.
We describe the experimental protocol and automated software for carrying out the data analysis and making images that encode the spatial distribution of nanostructural quantities of interest.
This supports a major goal in HT nanostructure characterization for situations with hundreds of measurements per hour and analysis times on the same order of magnitude as the measurement time~\cite{Potyrailo:2011jo}.  We refer to this approach generically as scanning nanostructure x-ray microscopy (SNXM).



The protocol is developed for screening spatially resolved PDF data and is modular in design.
This allows the protocol to be extended to a wide variety of high throughput experiments, such as in-situ synthesis experiments, as well as other experimental techniques.

\section{Experimental}\label{experiment}

\subsection{Sample preparation}
\begin{figure}[tbp]
	\centering
	\includegraphics[width=1\columnwidth]{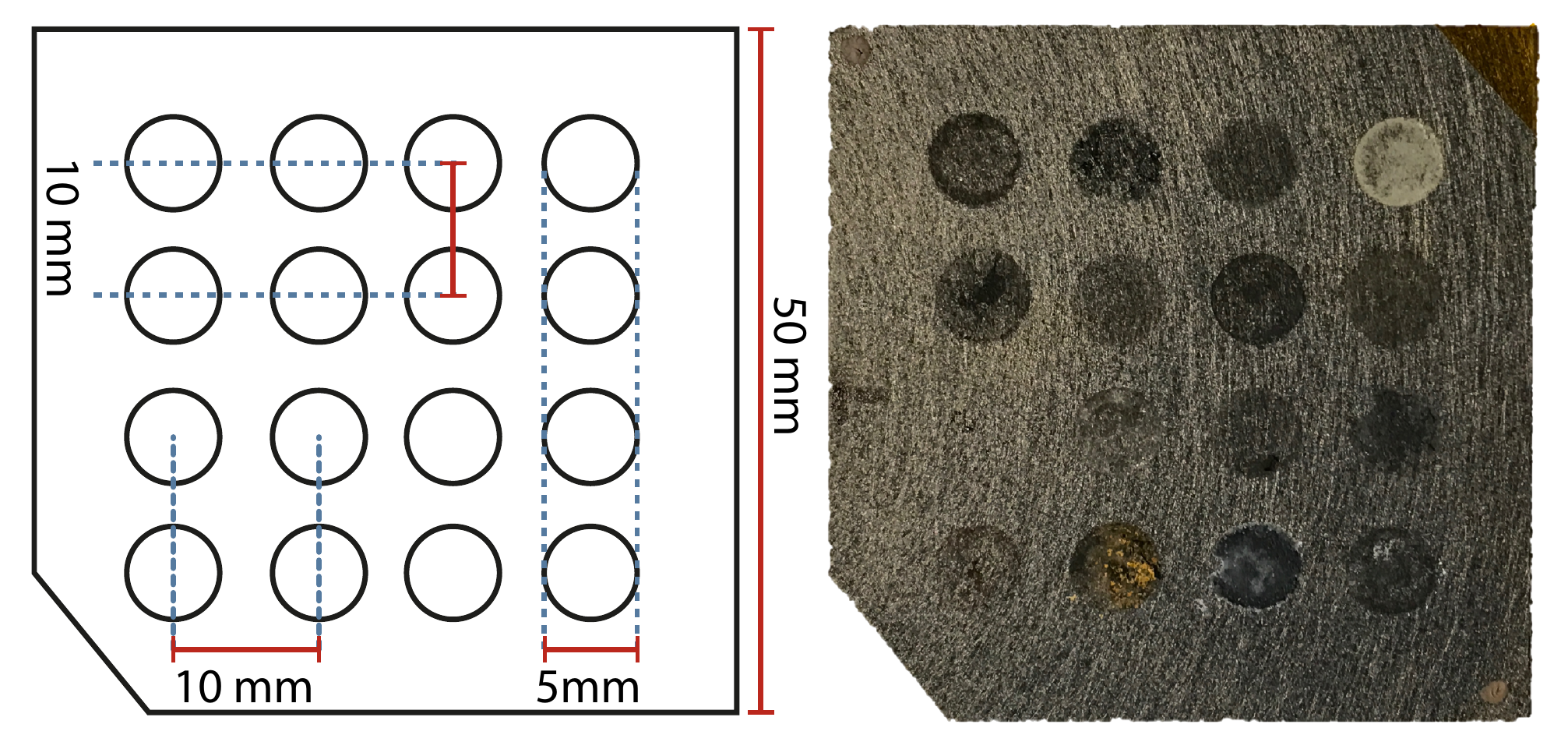}
	\caption{\label{fig:array_overview} A typical sample layout for combinatorial studies (left) and the tested array or catalytic material(right). A square piece of carbon paper was used as a substrate for the ink-jet printed material in a 4x4 configuration}
\end{figure}

The combinatorial catalyst library was deposited using a Pipetmax automated liquid handling system on semicrystalline carbon paper (Toray 120, from FuelCellStore) in a $4 \times 4$ grid giving 16 circular deposition sites (``wells'') 5~mm in diameter and with a center to center spacing of 10~mm~\cite{hittHighThroughputOptical2021}. Transition metal nitrate solutions at 0.1~M were used for deposition, except for the Au well where HAuCl$_{4}$ was used. The precursor solutions were mixed onto the carbon paper and reduced with excess hydrazine solution. The sample was then vacuum dried overnight in a 60~degree oven and washed with deionized water to create the differently alloyed metal samples on the substrate, as shown in Fig.~\ref{fig:array_overview}. The choice of chemicals, size, number of samples and pattern are programmable from the liquid handling system for future implementations of this protocol.

\subsection{Synchrotron x-ray measurements}
\begin{figure}[tbp]
	\centering
	\includegraphics[width=1\columnwidth]{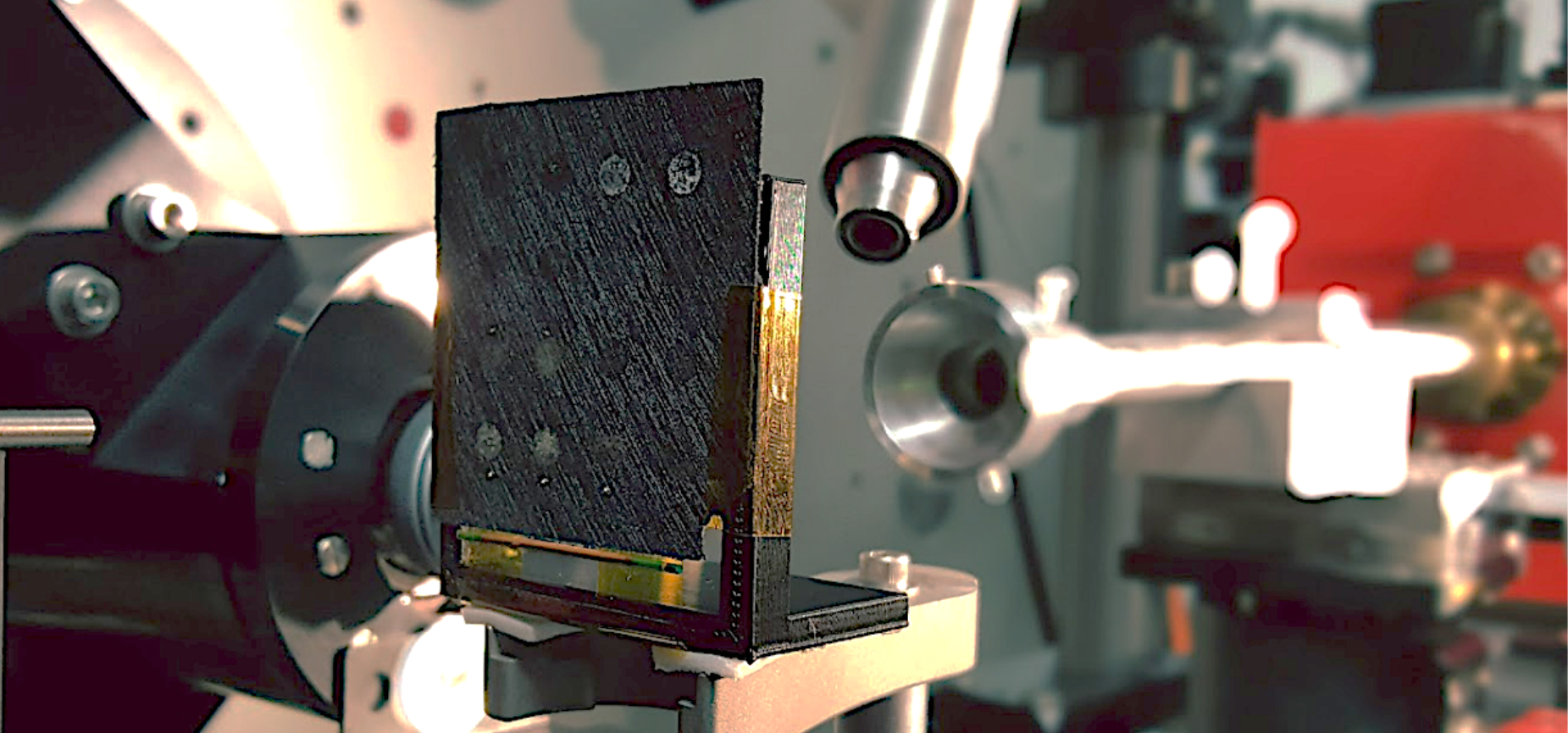}
	\caption{\label{fig:experimental_setup} The combinatorial library mounted on a 3D printed bracket in front of the x-ray beam. The array is mounted to the goniometer which allows measurement access to all deposition sites.}
\end{figure}

The experiments were carried out at the 28-ID-2 (XPD) beamline  at the National Synchrotron Light Source II, Brookhaven National Laboratory, using the normal incidence thin film PDF method~\cite{jense;ij15}.
The combinatorial array was mounted perpendicular to the x-ray beam direction using a 3D printed bracket.
The measurements were performed in a transmission geometry as shown in Fig.~\ref{fig:experimental_setup}.
The array was moved using goniometer motors in an $xy$-plane perpendicular to the incident beam direction, with a fixed sample to detector distance.
The 2D Perkin-Elmer detector was placed behind the sample at a distance of 203.4~mm, which gave an effective instrumental $Q$ range, where $Q=4\pi\sin\theta/\lambda$, of $0.12 \le Q\le 32$~\AA$^{-1}$.
The incident wavelength of the x-rays was $\lambda=0.183983$~\AA\ with a beam cross section at the sample of $250\times 300$~$\mu$m in the vertical and horizontal directions, respectively.

The sample wells are much larger than the beam, and the sample distribution within the wells is not uniform (see the Results section).
We therefore sought a measurement protocol that scanned over large areas of the sample in order to find both the best measurement conditions for sample determination, and also to assess the heterogeneity of the sample.
A zoomed-in measurement area of 9~mm by 15~mm was chosen over which the beam was scanned in a zig-zag linear array of points.
The chosen scan pattern encompassed two catalyst ``wells'' containing AuAg and AgCu nanocrystalline material, respectively, as shown in Fig.~\ref{fig:array_overview}.

A coarse alignment was done to set the position of the first measurement point by using a laser coaxially aligned with the incoming x-ray beam.
The zig-zag measurement pattern was then executed with a series
of 1~mm steps executed vertically, followed by a 1~mm horizontal offset, followed by 1~mm vertical steps in the opposite direction, repeated to cover the full measurement area.
Exposure time was selected, based on signal quality from a preliminary measurement on a nanoparticle spot, and set to $5$~s per point resulting in a measurement throughput of over $6000$ measurements per hour.

The sample-detector distance, $Q$ range and the geometric orientation of the detector were calibrated by measuring a crystalline Ni powder mounted on the same bracket that holds the sample chip prior to data collection from the sample itself.
The experimental geometry parameters were refined using the Fit2D program~\cite{hamme;esrf04}.
A mask was created to remove outlier pixels (dead pixels, hot pixels, and pixels shadowed by the beamstop) and applied to the 2D images from the measurement series before carrying out the azimuthal integration to a 1D diffraction pattern.

The carbon sheet produces a significant background signal in this experimental geometry, but the background signal can be subtracted from the data leaving only the structural information of the deposited material.
We found that background subtraction is not trivial for these samples and we developed a protocol for doing it that is described in the results section.
The total scattering structure function, $F(Q)$, was then obtained after standard corrections and normalizations of the data and Fourier transformed to obtain the PDF, using PDFgetX3~\cite{juhas;jac13} within xPDFsuite~\cite{yang;arxiv15}.
The maximum range of data used in the Fourier transform ($Q_{max}$) was chosen to be $21$~\AA$^{-1}$ in the current case, which was the best compromise between real-space resolution of the PDF signal, and noise.


\section{Results}

\subsection{Protocol automation software}
\label{sub:Protocol automation software}

The main goal of the protocol is to address the large number of measured data-points that are generated during high throughput experiments.
We have written a set of Python scripts that are intended to be highly flexible and customizable allowing for efficient data collection, curation, reduction and analysis.
The software is intended to be accessible and user friendly.
The code can be executed using IPython~\cite{Perez:2007wf} and Jupyter notebooks.

\begin{figure}[tbp]
	\centering
	\includegraphics[width=0.95\columnwidth]{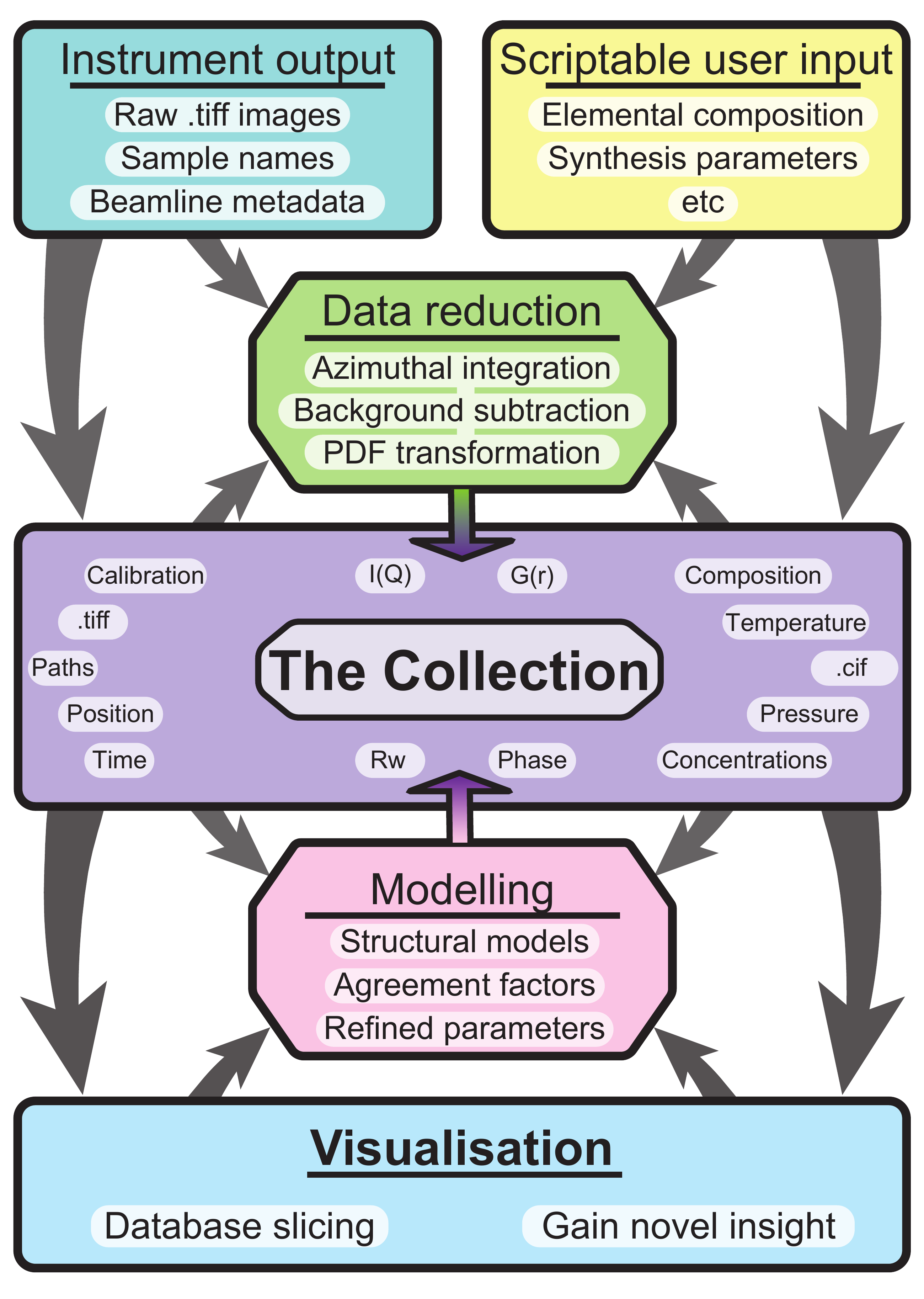}
	\caption{A flowchart illustrating the core of the mapPDF protocol and current implementation. Instrumental output 
  is combined with user created metadata to perform data reduction. Every step of the process is saved in the collection which can be sliced and visualized for screening and advanced analysis. 
  }
\label{fig:database_diagram}
\end{figure}

The overall approach builds a collection of information about the experiment, associating reduced data, user inputs based on prior knowledge of the material, and analysis results.
The collection can then be sliced and visualized easily by the user to interrogate and draw conclusions from the entire dataset.
A schematic of the overall layout is shown in Fig.~\ref{fig:database_diagram} showing all of the modules and the general workflow.

The data analysis protocol is currently optimized for the XPD beamline at NSLS-II.
After a measurement, the acquisition software at the beamline 
outputs a log file containing the metadata, such as motor positions, measurement times and unique identifiers for each diffraction image.

In the first step the protocol software interrogates the log file and converts each measurement entry into an event.
Each event then contains links to positional and other measurement metadata and the corresponding image files.
The main benefit of the approach is manageability of the contents of the collection which are easy to visualize for the user using standard python plotting packages such as matplotlib~\cite{Hunter:2007ux} in conventional 1D or heatmap plots by simple iteration and filtering of the corresponding keywords.
In addition we have prepared a few custom plotting functions that produce the figures presented. 

Any pre existing knowledge can be appended to the corresponding event entries using simple macros, such as python for loops and conditional statements. 
For example, we can add composition information based on our prior knowledge of the layout of the sample:
\begin{lstlisting}[language=Python]
  i = collection['x_motor'] < 1
  collection[i]['composition'] = 'AgCu'
\end{lstlisting}
This way, any useful information which is absent in the metadata can be added on a per entry basis.

Experimental geometry calibration information is obtained from a Ni standard material measured at the same time as the array (Fig.~\ref{fig:experimental_setup}).
Calibration parameters are used when the images are azimuthally integrated to one dimensional $I(Q)$ patterns using Fit2D.
The integrated patterns are then linked alongside the other information in the collection to the correct events as data arrays. 

\subsection{Background subtraction}
\label{Background subtraction}

The tfPDF measurement requires careful subtraction of the substrate scattering because the substrate signal (background) was significant compared to the small signal from the deposited nanoparticles. 

Background images were acquired from a sample region with no nanomaterial and integrated to $I(Q)$ in the same way as the images containing the material.
A different background measurement can be assigned to each entry in the collection. This is particularly useful when substrate properties vary as a function of position and a background measurement in proximity to the material of interest is optimal for signal extraction.  In the present case a single background dataset was collected from the center of the array and assigned to all images.

The background subtraction is performed after interpolating the background dataset onto the $Q$ grid of the target pattern.
In most cases a scaling factor of 1 is used for all backgrounds in a dataset, but a global scale factor may be defined by the user if needed. 
Additionally, a utility function has been provided to optimize the background subtraction per entry in the collection, by minimizing the difference between sample and background signal intensity, over a user defined $Q$-range~\cite{jacqu;nc13}.
The background subtracted diffraction patterns are then appended to the collection as data arrays. 

\subsection{PDF transformation and model fitting}
\label{PDF transformation and model fitting}

\begin{figure}[tbp]
	\centering
	\includegraphics[width=1\columnwidth]{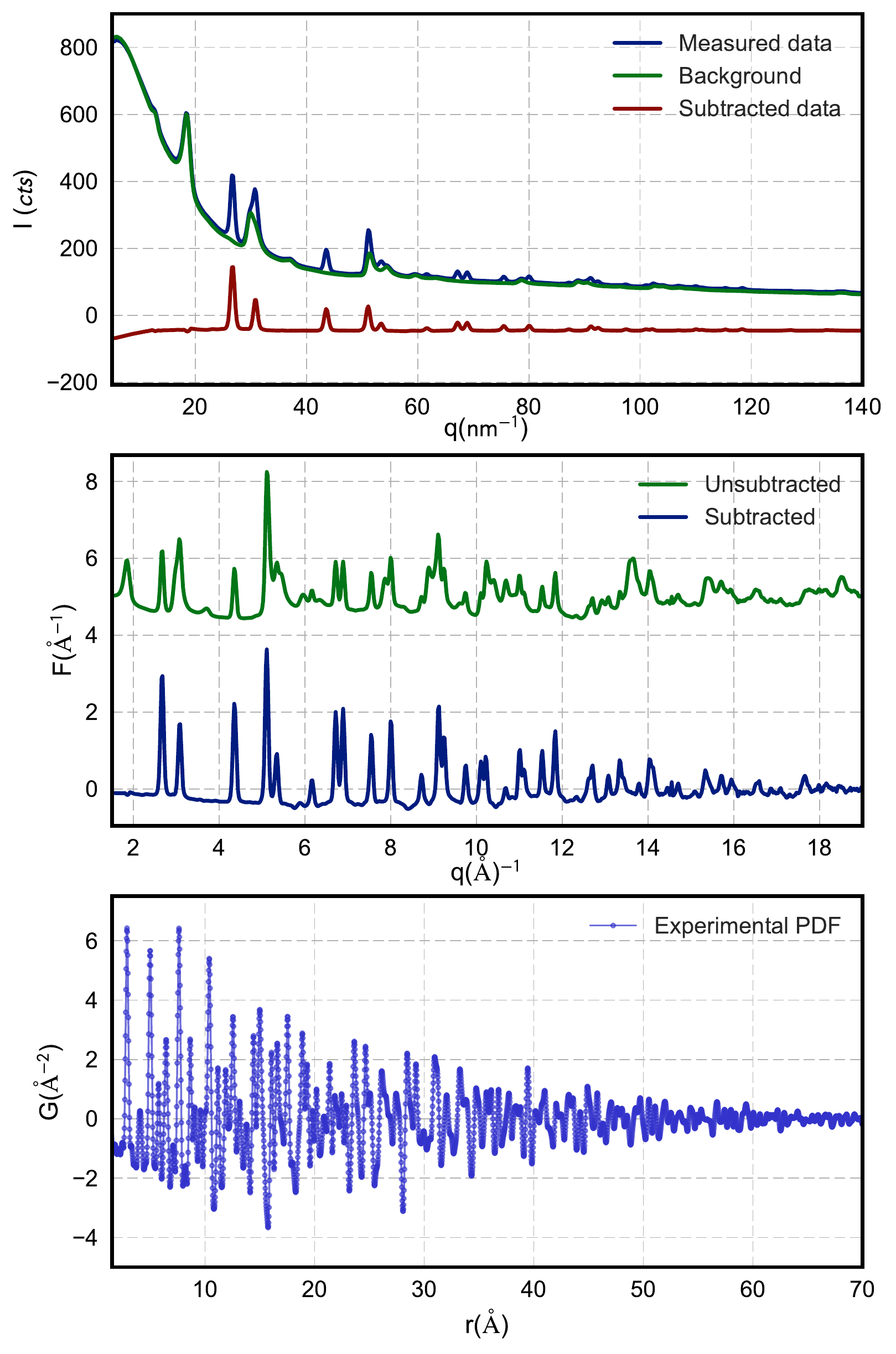}
	\caption{\label{fig:processing_example} An example of data processing from a single event in the collection.
  The background signal is subtracted following normalization in order to better resolve scattered intensity from the nanoparticle sample.} 
\end{figure}

The background subtracted $I(Q)$ data is Fourier transformed to the PDF using PDFgetX3~\cite{juhas;jac13} using parameters such as Q$_{max}$ and elemental compositions that are stored in the collection. 
The output PDF data, $G(r)$, is again appended to the main collection. A representative example of data at each step of the process is shown in Fig.~\ref{fig:processing_example}.
These transformation steps can be performed on all database entries or a subset.

\begin{figure}[tbp]
	\centering
	\includegraphics[width=1\columnwidth]{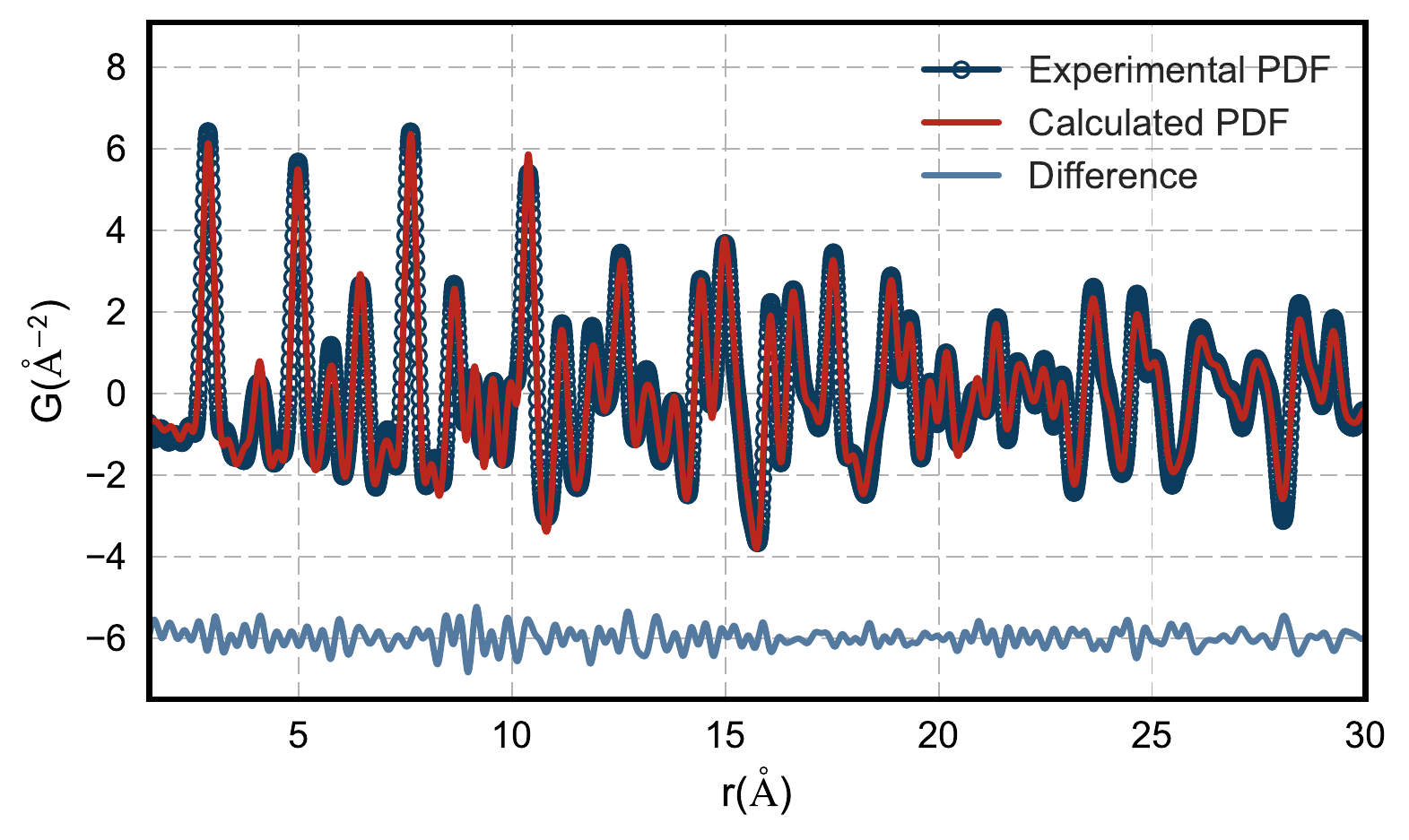}
	\caption{\label{fig:fit_example} An example of a single PDF fit using a bimetallic FCC model to one of the datafiles 
  in the collection. Blue illustrate
  the experimental data. The refinement score (\rw ) is 12.3$\%$.}
\end{figure}

In the combinatorial array experiment presented, each well contained different metallic nanoparticles.
We used an fcc model 
to refine the experimental PDFs and extract structural parameters for each event in the collection.
The PDFs, relevant metadata entries and initial guesses for the structural parameters are fed into the model to perform
structural refinement using the \textsc{Diffpy-CMI}~\cite{juhas;aca15} - Complex modelling Infrastructure software available at \url{diffpy.org}. 
A representative example from the combinatorial array experiment may be seen in Fig.~\ref{fig:fit_example}.
The primary parameters of interest from the output of the refinement for this dataset, namely the crystallite size and lattice parameter, and weighted agreement factor $R_w$, are associated with the correct event in the collection,
as shown in Fig.~\ref{fig:database_diagram}.

\subsection{Visualizing spatially resolved data}
\label{sub:visualizing_combinatorial_data}

\begin{figure}[tbp]
  \centering
  \subfigure[]{
    \includegraphics[width=1\columnwidth]{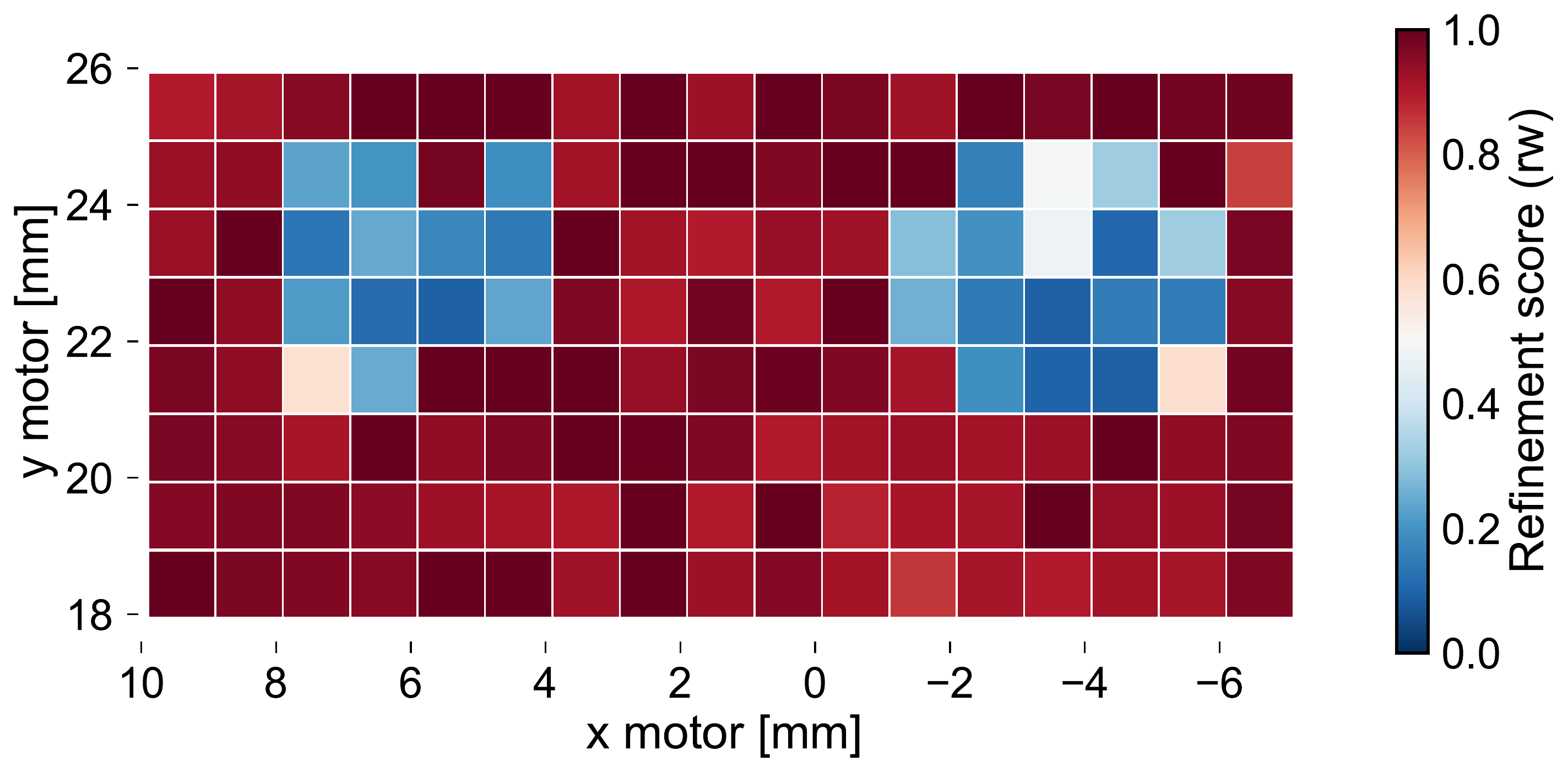}
    \label{fig:gridscan_rw}
  }
  \subfigure[]{
    \includegraphics[width=1\columnwidth]{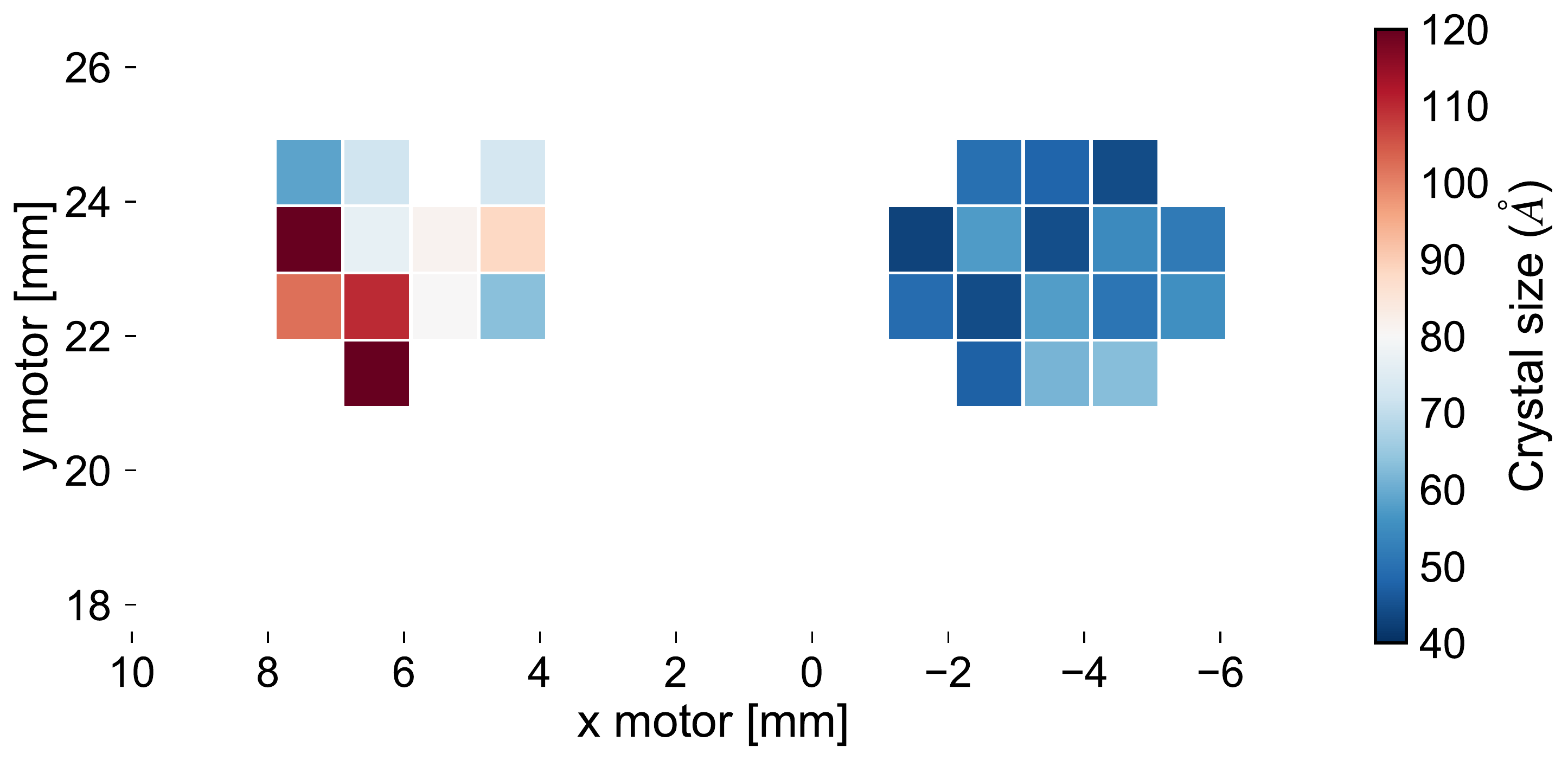}
    \label{fig:psizegridscan}
  }
  \caption{\ref{fig:gridscan_rw} Map of refinement scores
  vs. position for the array. Red squares indicate a measurement area with a poor refinement score, while blue squares indicate areas with good refinement scores, and thus presence of the fcc phase.
  There are two distinct regions with nanomaterial surrounded by measurements of nothing but the background.
  The refinement scores for this dataset are highly correlated with signal to noise ratio and give an indirect metric for the amount of material in a given area.
  \ref{fig:psizegridscan} Map of particle size vs. position on the array filtered to only display good model refinement scores. The colorscale indicates spherical particle diameter parameter
  from smaller (blue) to larger (red) crystallite size estimates.
  The figures are generated using simple conditional statements to slice the collection.}
  \label{fig:gridscan_parameters}
\end{figure}

Good visualization tools are essential for HT experiments.
The approach outlined above results in a comprehensive collection of measured data and data analysis results.
Presenting this data in a manageable way is usually a major challenge.
The main philosophy we have taken is to make spatial maps of scalar quantities 
that are associated with some aspect of the components in the collection, for example, goodness of fit or lattice parameter.
Figure~\ref{fig:gridscan_parameters} illustrates a usecase where position of the quantity on the plot corresponds to the physical position on the chip where the data were measured, as viewed along the direction of travel of the x-rays.
Fig.~\ref{fig:gridscan_rw} shows the $R_w$ from fits of the fcc model to the background subtracted data from our array of catalytic material as a function of position on the array.
The plot can be generated from a complete collection using a simple plotting function:

\begin{lstlisting}[language=Python]
    slice_2D('x_motor','y_motor','rw')
\end{lstlisting}
this function loops over the collection, extracts the parameter of interest, sets the correct boundaries, and sets a colorscale for the false-color plot.
The color of the squares indicates the model fit quality at the measured position, where dark red indicates minimal or no agreement to the candidate structure, and dark blue for good agreement.
After the background subtraction step, the areas where all of the signal is from the substrate would contain nothing but noise.
Since we are fitting the fcc model, these regions will result in poor \rw\ values and good fits are an indication of where the catalytic material is located, and how much is there.
We can then return
to the locations exhibiting a better fit, which contain signal from the material of interest, to do more careful structural analyses.

In a similar fashion to the figure above, it is possible to generate maps of any quantity in the collection with multiple filters
by using simple python {\tt for} loops, conditional statements and built in matplotlib functions like the one presented in Fig.~\ref{fig:psizegridscan}:
\begin{lstlisting}[language=Python]
i = d['rw']<0.3
plt.scatter(d['x'][i], d['y'][i], c=d['psize'][i]
\end{lstlisting}

The code snippet above generates the spatial map of nanoparticle size vs. position refined from the fcc model after filtering for an acceptable \rw\ threshold.
From the figure it becomes clear that the particle size distributions differ within the wells with the AgAu well being much more uniform and smaller on average.

\subsection{Software flexibility, modularity and availability}

The software that implements the protocol can be divided into several key parts as presented in Fig.~\ref{fig:database_diagram}.  These are initial data treatment, transformation of the data and model fitting and refinement. 
Because of the modularity, all three can be modified, replaced or omitted by the user depending on the use-case and user preferences allowing the user to easily build a bespoke analysis for their data.

Although originally intended for tracking positional information about the sample, the protocol can be extended to keeping track of any scalar quantity and has been found extremely useful for time-series datasets.
Structural parameter evolution as a function of time, instead of being a function of motor positions, can be visualized using the protocol software and helps streamline systematic analyses of large {\it in situ} datasets, for example.
The methodology is currently being extended for studying nanocluster formation in a wet synthesis environment measured at the P.02 beamline at PetraIII, Hamburg, Germany.

The latest source code is available open source with a BSD license on GitHub as part of the diffpy organization at \url{https://github.com/diffpy/mappdf} as well as an example dataset used to generate the figures above.

\section{Conclusion}

An analysis 
protocol and a set of scripts for treating a wide variety of combinatorial high-throughput materials characterization data is presented. 
The protocol software is flexible and can be modified and expanded by the user.
An example of a combinatorial catalyst library analyzed using the PDF technique has been demonstrated, highlighting the power of the approach.
The ability keep track of and analyze large volumes of data and additionally parameterize 
the dataset to allow for quick analysis that is necessary for high throughput experiments.

\section{Acknowledgements}
PDF methodology developments were funded by the DOE Office of Science by Brookhaven National Laboratory under Contract No. DE-SC0012704. AK acknowledges funding by the Innovation Fund Denmark (Green Chemistry for Advanced Materials 4107-00008B-GCAM). Sample preparation was supported by the Office of Basic Energy Sciences, Division
    of Chemical Sciences, Geosciences, and Energy Bioscience, Department of Energy
    under contract (SC-0019781). X-ray PDF measurements were conducted on beamline 28-ID-2 of the National Synchrotron Light Source II, a U.S. Department of Energy (DOE) Office of Science User Facility operated for the DOE Office of Science by Brookhaven National Laboratory under Contract No. DE-SC0012704.

\bibliography{billinge-group,abb-billinge-group,everyone,map_PDF}
\bibliographystyle{aip_simon}

\end{document}